# Large-scale molecular dynamics investigation of geometrical features in nanoporous Si

Laura de Sousa Oliveira[*] and Neophytos Neophytou

School of Engineering, University of Warwick, Coventry, CV4 7AL, UK

[*] L.de-Sousa-Oliveira@warwick.ac.uk

## Abstract

Nanoporous materials are of broad interest for various applications, in particular advanced thermoelectric materials. The introduction of nanoscale porosity, even at modest levels, has been known to drastically reduce a material's thermal conductivity, in some cases even below its amorphous limit, thereby significantly increasing its thermoelectric figure of merit, $ZT$. The details of the important attributes that drive these large reductions, however, are not yet clear. In this work, we employ large-scale equilibrium molecular dynamics to perform an exhaustive atomistic-scale investigation of the effect of porosity on thermal transport in nanoporous bulk silicon. Thermal transport is computed for over 50 different geometries, spanning a large number of geometrical degrees of freedom, such as cylindrical pores and voids, different porosities, diameters, neck sizes, pore/void numbers, and surface-to-volume ratios, placed in ordered fashion, or fully disordered. We thus quantify and compare the most important parameters that determine the thermal conductivity reductions in nanoporous materials. Ultimately, we find that, even at the nanoscale, the effect of merely reducing the *line-of-sight* of phonons, i.e. the clear pathways that phonons can utilize during transport, plays the most crucial role in reducing the thermal conductivity in nanoporous materials, beyond other metrics such as porosity and surface/boundary scattering.

# I. Introduction

Nanoporous materials, particularly nanoporous Si-based materials, have received significant attention in the past three decades for a variety of applications [1], including photonic [2], optoelectronic [3] and microelectronic devices [4], data storage [5], functionalized sensors and filters for chemical/biological applications [6], and more recently as thermoelectric materials [7-9]. With regards to thermoelectric materials, good performance is determined by low thermal conductivities. The thermoelectric performance is quantified by the dimensionless figure of merit, $ZT$, defined as $ZT = (S^2\sigma)/\kappa$, where $T$ is the temperature, $S$ is the Seebeck coefficient, $\sigma$ is the electrical conductivity, and $\kappa$ is the thermal conductivity.

Interest in Si-based nanoporous materials for thermoelectric applications is largely due to their two orders of magnitude lower thermal conductivity compared to bulk. Such reductions have been observed over the last few years in several other Si-based nanostructures as well, e.g. in rough Si nanowires [10,11], thin films [12,13], and Si-based alloys and superlattices [14]. Recent works have also shown that the room temperature thermal conductivity of Si-based nanoporous materials can even be reduced beyond the materials' amorphous limit [9,15,16]. These observations are attributed to strong phonon-boundary scattering and make Si-based nanoporous materials excellent candidates for next-generation thermoelectric applications, as $ZT$ is inversely proportional to the material's thermal conductivity. Specifically, for nanoporous materials the thermal conductivity reported values are in the range of 1–2 W m$^{-1}$ K$^{-1}$ (drastic reduction compared to the bulk, $\kappa_{\text{bulk}} \sim 150$ W m$^{-1}$ K$^{-1}$), with a $ZT$ of ~0.4 [9] (drastic increase compared to bulk Si, $ZT_{\text{bulk}} \sim 0.01$). Moreover, several experimental and theoretical works indicate that porosity can also be designed in such a way as not to degrade the thermoelectric power factor, $S^2\sigma$, and possibly even increase it in some cases [9,15,17-19]. Since the mean-free-paths of electrons and phonons are different, carefully designed porosity allows for the reduction of thermal conductivity, while maintaining a significant degree of crystallinity, and thus retaining high power factors [20]. To optimize the porosity in this way, we need to understand in detail the mechanisms by which geometrical variations in porosity reduce the thermal conductivity.

The effect of pore/void geometry and arrangement on thermal conductivity is complex. Several computational studies have addressed various geometrical aspects of porosity, e.g. variations in pore surface area [21-24], pore numbers [23,25], sizes [21,23,25,26], shapes [23,27], distances [21,22,26], boundary roughness [24,25,27,28], and amorphicity [22,29]. Molecular dynamics (MD) [21,22,25,29-31], often coupled with lattice dynamics [22,32], and Monte Carlo (MC) Boltzmann transport equation (BTE) solvers [12,23,24,33] are the most commonly used computational approaches to investigate thermal transport on nanoporous materials.

However, while studies to date have reported on different individual phenomena, a clear and complete understanding of the physical mechanisms that degrade the thermal conductivity, in the combined context of porosity, disorder, pore placement, pore clustering, etc., has not yet been reached. All those attributes will of course reduce the thermal conductivity, but the importance of each mechanism is not clear yet. This is also evident from the conflicting results that one encounters in the literature, especially with regards to the relative strength of phonon scattering mechanisms, such as the effect of surface area versus the distance between the pores, the pore size, porosity, pore misalignment and randomization (which could produce non-propagating diffuse phonon modes [21,22]), etc. For this, large-scale atomistic simulations of embedded nanopores with relatively large domains are needed, in order to capture the effect of nano-scaled pores within 3D domains.

In this work, we employ molecular dynamics simulations to perform an exhaustive investigation of the effect of thermal conductivity in nanoporous Si. For this we have computed the thermal conductivity of over 50 different nanoporous structures (each simulated for at least 10 initial configurations) using the Green–Kubo approach within equilibrium molecular dynamics (EMD). To capture a wide array of geometrical degrees of freedom and phonon scattering lengths we employ relatively very large (for MD) domain sizes, of up to 108 nm in length, with typically ~160,000 atoms. We consider cylindrical pores as well as voids (spherical pores — referred to as *voids* in the paper), and examine a series of geometrical degrees of freedom, e.g. porosity, diameter, neck size (i.e. distance between pores), pore number, and surface-to-volume ratio. In order to identify the most important geometrical features that are responsible for the reduction in the thermal conductivity we investigate multiple distributions, including uniform, staggered, clustered and fully randomized cylindrical and spherical pore distributions. While all of these have their degrading influence on thermal conductivity, we conclude that it is the reduction in the *line-of-sight* of phonons which has the strongest effect. In other words, as in a particle-like phonon picture, it is the blocking of phonon trajectories in the transport direction (which shows up in staggered and disordered configurations), that has the most marked effect on the thermal conductivity. The paper is organized as follows: Section II describes the computational *Approach*, Section III describes the *Results and Discussion*, and finally we conclude in Section IV.

## II. Approach

The focus of this study is to elucidate the main features that affect the flow of phonons as they propagate through nanoporous geometries with various pore/void arrangements. The geometries we simulate range from uniform distributions of pores and voids with various sizes and numbers of pores/voids, to staggered, clustered, simultaneously staggered and clustered, and randomly distributed (and sized) pores/voids, as illustrated in Fig. 1. We employ classical equilibrium molecular dynamics (EMD) to perform thermal transport calculations. Molecular dynamics (MD) allows for substantially larger system sizes than first principles approaches, and is therefore the most commonly used atomistic approach for thermal transport. It also has the advantage of capturing the anharmonicity of interatomic interactions which is implicit in MD through the choice of the potential. On the other hand, particle-based real-space approaches (such as Monte Carlo) often require simplifying assumptions about the nature of relaxation times, or that predictions be fitted to experimental or theoretical data. The use of molecular dynamics bypasses the need to estimate and instead allows us to investigate scattering mechanisms which are implicit in the choice of potential, as well as the atomistic details of the physical structure. By not treating phonons as particles, MD can also capture wave effects, such as coherence/decoherence, and merge the phonon nature of waves and particles. (Note that purely real-space wave methods, such as implementations of the non-equilibrium Green's function method for phonons, commonly do not include anharmonicity, which overemphasizes the wave nature [34].) In EMD, the use of periodic boundary conditions accommodates mean-free-paths (MFP) larger than the size of the simulation cell, and provides the complete thermal conductivity tensor in a single simulation. In this work we simulate systems of ~108 nm in length with a 5.43 $nm^2$ cross section. The simulated geometries have approximately 160,000 atoms, which is above the ~64,000 atom thermal conductivity convergence threshold suggested in Ref. [35] for pristine single crystalline Si. A plot of system size convergence is offered in the *Supplemental Material*. The selected length is also sufficiently large to capture the influence of pores in a wide range of configurations.

To extract the thermal conductivity we use the Green–Kubo method [36,37], a widely used and well-established equilibrium molecular dynamics (EMD) approach. The Green–Kubo formalism relies on the assumption that the same mechanisms, or processes, by which a system responds to a stimulus or perturbation (e.g. temperature gradient) are also responsible for its response to local fluctuations (e.g. in the heat flux) in equilibrium. Mathematically, this means the first element in the thermal conductivity tensor, corresponding to thermal transport along the *x*-direction, can be calculated as:

$$\kappa_{xx} = \frac{V}{k_B T^2} \int_0^\infty <J_{xx}(t)J_{xx}(t+\tau)> d\tau, \tag{1}$$

where $V$ is the volume of the simulated region, $T$ is the temperature, and $J_{xx}$ is the first element of the heat-flux (or heat current) tensor. The expression $<\boldsymbol{J}(t)\boldsymbol{J}(t+\tau)>$ is the non-normalized heat-current autocorrelation function (HCACF), $t$ designates the simulation time, and $\tau$ the autocorrelation time. The HCACF can be computed as the inverse Fourier transform of the same transform of the heat-current (as a function of simulation time) multiplied by its complex conjugate, or numerically as:

$$<\boldsymbol{J}(t)\boldsymbol{J}(t+\tau)> \equiv \sum_{n=0}^{N-m} \frac{\boldsymbol{J}_n \boldsymbol{J}_{n+m}}{N-m}, \tag{2}$$

where $\boldsymbol{J}_n$ is the value of $\boldsymbol{J}$ at the $n^{th}$ time step, and $\boldsymbol{J}_{n+m}$ is $\boldsymbol{J}$ at the $(n+m)^{th}$ time step, for $n$ = 0, 1, 2, …, N and $m$ = 0, 1, 2, …., M. Here, N and M are the maximum number of steps in the simulation and in the HCACF, respectively. The random-walk nature of the error in the autocorrelation function further means that the error in the estimated thermal conductivity, which is a function of the integral of the autocorrelation function, grows over time. In practice, the HCACF is therefore truncated at an earlier time. Oftentimes, a compromise has to be made between an earlier cut-off that reduces the error in the thermal conductivity, but could possibly neglect slower relaxation processes, or vice-versa. For systems which have large phonon MFPs, the HCACF converges to zero at a slower rate, which results in a higher error, and thus, variability in thermal conductivity for the different trajectory simulations. We have opted to truncate the HCACF at 150 ps for most systems, with the exception of the pristine geometry for which the cut-off was selected at 500 ps. A more thorough discussion on the topic of error mitigation can be found in Ref. [38], and a discussion/justification on cut-off selection, can be found in the *Supplemental Material*. Moreover, the results were averaged for at least 10 sets of simulations (sometimes more) to mitigate the large uncertainty in the Green–Kubo approach. Simulations are ~ 5.4 x 5.4 x 108.6 nm³, corresponding to a 10 x 10 x 200 supercell of the 8 atom Si unit cell (see inset in Fig. 1a), i.e. ~160,000 atoms, with the *x*-axis corresponding to the [1 0 0] direction.

Simulations were performed with the large-scale molecular dynamics software LAMMPS [39], using the Stillinger–Weber (SW) potential [40]. This potential has been widely used to model heat transfer in silicon [35,41], and while it is known to overestimate thermal conductivity [42], it has successfully been used to describe elastic constants and thermal expansion coefficients and offers a reasonable match for phonon dispersion relations [43], especially for acoustic phonons. As is common practice, we report on the fractional change in thermal conductivity between the porous systems compared to the pristine system, $\kappa_{porous}/\kappa_0$. In this work, we consider both systems with spherical and cylindrical holes, which are henceforth designated as voids and pores, respectively. Thus, voids (spherical holes) are empty spheres within the 3D domain, while pores (cylindrical holes) are empty cylindrical porous regions that are 'etched' from the top all the way to the bottom of the material. In the presence of voids, in a 3D domain, phonons can flow around them in '3D' pathways. In the cylinder-like pore case, phonons will have to essentially flow around them in '2D' pathways. The pores/voids are created by deleting atoms in the minimized pristine geometry. After the pores are introduced, each system in a set (i.e. for a given geometry) is then independently given a thermal energy equivalent to ~ 300 K, so that each system in a set has its own initial configuration. This is done by generating an ensemble of velocities with a Gaussian distribution using a different random seed (and thus rescaling the velocity) each time. The systems are then allowed to equilibrate to room temperature within an isothermal, isobaric ensemble (NPT) which allows them to thermally expand. In the next step, the systems are equilibrated in the microcanonical ensemble (NVE) for an additional 125 ps, with a 0.5 fs interval (as in the previous step). Finally, the simulations are resumed in the microcanonical ensemble at a 2 fs time step for an additional 10 ns during which time we record the heat-flux for the HCACF calculation. Ultimately the HCACF is computed for a 20 fs time step, having confirmed that the choice of time step in no way changes our results compared to smaller steps.

## III. Results and Discussion

A wide number of geometries (upwards of 50, each simulated at least 10 times) were considered in this study. A table of the geometries and corresponding thermal conductivities can be found in the *Supplemental Material*. We compute porosity as the fraction of the number of atoms removed from a porous geometry to that of the pristine system (equivalently, porosity is the fractional volume of the pores/voids to the total volume of the system). As explained in the *Approach* section, the thermal conductivity of the porous systems is normalized to the pristine channel thermal conductivity, $\kappa_0$, which is estimated at 347.8 ± 34.7 W m$^{-1}$ K$^{-1}$ with the Stillinger–Webber potential. Note that it is standard process to show normalized results in this kind of studies, since different potentials, each with their own advantages, provide different values for the thermal conductivity [22,29,41]. Unlike the porous systems, for which the cut-off was selected at 150 ps, the pristine system thermal conductivity is cut-off at 500 ps. This is because large mean-free-paths, in the order of several hundreds of nanometers, are present in pristine silicon, but cease to be prevalent in porous materials, especially as porosity increases since the mean-free-paths are then mostly determined by the pores (see the *Supplemental Material* for a more in-depth discussion on the choice of cut-offs).

The thermal conductivity versus porosity for all of our simulated geometries is shown in Fig. 1a. In this figure we include some characteristic structure families as depicted in Fig. 1b, i.e. geometries in which the pores are distributed uniformly, staggered, clustered, simultaneously staggered and clustered, randomized in terms of position, and randomized in terms of position and size (notice that the colouring of the geometries corresponds to the colouring of the data points in Fig. 1a). Following the arrows in Fig. 1b, our intent is to explore the different effects due to geometry, but also the incremental influence of each feature. Porosity is known to have a significant influence on the thermal conductivity, and this is what we also observe in our simulation results, where even for small porosities, a drastic reduction in $\kappa$ is observed. This agrees with other works in the literature, for example, He *et* al. [22] find a reduction of about an order of magnitude in $\kappa$ at 7 % porosity for a 20 nm thin-film with cylindrical pores, in good agreement with experiments [12]. In Fig. 1 we observe that the simulated systems approach a 10-fold decrease in thermal conductivity at porosities as low as 2%. This is a large decrease, but it is well known that the thermal conductivity in nanoporous materials deviates significantly from macroscale porous materials, which follow the Eucken model in which $\kappa$ is reduced linearly with porosity [24,29]. We note, however, that in our simulations, in most cases pores/voids are closely packed perpendicular to the transport direction, clearly forming planes/surfaces of high thermal resistance where the pores/voids are concentrated (through periodic boundary conditions in the *y*-direction), which will have a more drastic effect on $\kappa$ along the *x*-direction. As a result, the systems studied in this work are anisotropic. These plane barriers which, even at low numbers of pores/voids, hinder phonon propagation along the *x*-direction, and thus the sharp drop in Fig. 1a for the systems with very low porosities, < 0.5%. However, this is done on purpose, since our intent is not to provide thermal conductivity predictions, but investigate the most important details that determine $\kappa$ reductions in porous materials. Below, we proceed to a more detailed analysis of the results of all simulations, beginning with the effect of void/pore surface area on thermal conductivity.

<u>Influence of void surface-to-volume ratio:</u> We first consider the effect of surface area in systems with voids, by computing the surface-to-volume ratio, $\rho$, defined as the total surface of the pores/voids to the total volume of the geometry (including the empty voids/pores) [22,44]. In Fig. 2a multiple systems with voids with equivalent porosity (~0.45%), but different surface areas, are depicted. To increase the surface area while maintaining the porosity constant, we increase the number of voids while reducing their radius, as indicated in the schematics of Fig. 2a. In this way, at the same porosity, we consider the effect of increasing surface area (Fig. 2b), but also the resultant effect of increasing the number of voids (Fig. 2c), both of which influence phonon scattering. Smaller voids are less effective scatterers compared to larger voids, but the larger number of voids/pores decreases the distance between scattering events, thereby reducing the overall phonon MFP. For systems of similar porosity (~0.45%), in Fig. 2a we plot the thermal conductivity versus the surface-to-volume ratio, $\rho$, and in Fig. 2c versus the number of scatterers. The coloring of the data points corresponds to the geometry coloring in Fig. 2a: this includes four systems with $\phi \sim 0.45\%$, and a fifth geometry with a lower $\phi = 0.34\%$ (in dark blue), but with a surface area comparable to the other four systems. As expected, we find that there is a clear trend with $\kappa$ decreasing as the surface area increases, and as the number of scatterers increase as well. More specifically, the results in Fig. 2b suggest that by doubling the surface area, there is an approximate 19% decrease in thermal conductivity (although we note that due to the nature of the MD simulations a larger uncertainty is associated with lower porosity results, as quantified by the reported error). The results in Fig. 2 indicate that surface area and number of defects can be more useful metrics in determining $\kappa$ compared to porosity in these nanoscale systems. Indeed, the fifth geometry we consider (in dark blue, with the lower $\phi = 0.34\%$), has lower $\kappa/\kappa_0$ compared to the light-blue larger porosity structure, indicating that the surface area could be a stronger predictor in determining $\kappa$ rather than the porosity. Other works [21,22,29], have also reported on the effect of pores/voids' surface area on thermal conductivity. Lee *et* al. [21] alternately varied the pore diameter and the distance between pores, with their results indicating in addition that surface area has a greater impact on thermal transport than the distance between pores.

<u>Influence of pore surface-to-volume ratio and geometry asymmetry:</u> We next consider the same effect of surface area/scatterer number, but in systems with cylindrical pores (see Fig.

3). Two sets of systems are compared, one with small porosity $\phi = 1.2\%$ (blue and green), and one with higher porosity $\phi = 10\%$ (orange and cyan), shown in Figs. 3a and 3b, respectively. Here, in order to investigate the effect that the distance between pores exerts on the thermal conductivity, we consider also transport along the *y*- and *z*-directions in addition to transport along the *x*-direction. We then have the following situations: i) while travelling along *x* (left-to-right), phonons directly scatter on 'surfaces' of closely packed pores, ii) while travelling along *y* (bottom-to-top), phonons scatter on pores less closely packed (blue, green, and orange), and iii) while travelling along *z* (into the page), phonons travel along the surfaces of pores. Since, for each geometry, the surface area and total number of scatterers per volume is maintained in all directions, differences between transport in *x*, *y,* and *z* are therefore due to the anisotropic geometry and the spacing between the pores that the phonons encounter. In Fig. 3c we compare the thermal conductivity of these systems versus their surface-to-volume ratio. For each of the systems in Fig. 3(a), clearly, transport along z (hexagons) has the least disruption in k, followed by transport along y (rhombus), and then along x (for the geometries in blue, green and orange), in which case phonons directly traverse the pore 'wall' regions. Comparing the blue and green systems which have the same porosity and similar surface areas, we observe that along *x* and *y* (square and rhombus blue/green symbols in Fig. 3c), they have comparable thermal conductivities. Phonons in the blue system encounter fewer, but larger (more effective) scatterers. Phonons in the green system encounter more scatterers, more closely packed along *y* (and *x)*, but smaller, which makes each individual pore less effective [45]. Finally, the thermal conductivity is similar in the two systems. In the *z*-direction (hexagons), however, since the phonons travel along the pore surfaces and interact with them continuously, the larger pore number (and surface area) in the green system reduces thermal conductivity more strongly.

In the case of larger porosity systems (orange and cyan), we have many more pores and the structures seem more isotropic. Within each geometry, the differences between transport directions are not as severe; in fact, for the geometry in cyan, transport along *x* is symmetric to transport along *y* (the pores are equidistant in both directions). Comparing the orange to the cyan systems, we find the following: i) Along *x,* in the orange system, phonons encounter dense 'surfaces' of pores with a smaller distance between them. In the cyan case, phonons encounter less 'dense' surfaces, with larger distance between the pores perpendicular to transport, but increased number of pores. The larger number of pores compensates the smaller diameters, but still, the thermal conductivity of the cyan system, with more space for phonons to go through the pores, is slightly higher by 18% (compared to the orange system). ii) On the other hand, along *y* the thermal conductivity in the orange system is higher compared to the cyan by ~62%. Phonons in the orange system have larger areas, 'completely' free of pores to travel, whereas phonons in the cyan system encounter more surfaces in their path. iii) Along *z* (into the page, hexagonal symbols), in a similar manner, the orange system supports larger uninterrupted areas, while the cyan system forces phonons to travel more closely along the pore surfaces, and thus thermal conductivity is lower in the cyan system by ~48%.

From these observations, we can clearly state not only that surface area is an important parameter that influences thermal conductivity (given observations for transport along *z* and to some extent *y*), but it also makes a large difference how closely packed the pores are with respect to phonon transport (given observations for transport along *x* and to some extent *y*). Transport in the *x*-direction (square symbols) is degraded the most when phonons encounter a dense 'surface' of pores, with small pathways for phonons to pass through, which introduces large thermal resistance (compared to encountering more pores and more surfaces). This effect is referred to in the literature as reducing the phonons' *line-of-sight*, a term indicating whether particle-like phonons have a long direct path from one side of the material to the other, without interruptions [46,47]. Concerning transport perpendicular to the pores, *line-of-sight* has a stronger influence compared to the number of pores and surface area in systems with closely packed pores.

Along *z*, phonons travel along the pores, and thus interact and are affected by the surface area continuously as they propagate. The closer the surface is to the phonons, the more its influence is expected to be. When pores are further apart, even if their diameter is larger (blue/orange), their influence is reduced compared to closely packed pores with phonons travelling along them (green/cyan). Thus, it is not only if phonons scatter on the pores directly themselves, but surface scattering when travelling along pores also affects the thermal conductivity, in a similar manner to how surfaces reduce electron mobility [48]. Indeed, nanowires are known to significantly reduce thermal transport due to scattering along the boundaries [31,45,49].

<u>Influence of pore asymmetry in transport direction:</u> In Fig. 4, we generalize some of the observations from Fig. 3 by simulating and plotting the directional thermal conductivity of multiple geometries of uniformly distributed pores (as in the inset of Fig. 4) as a function of porosity. Transport in the *x*-direction (square symbols) results in the lowest thermal conductivity compared to transport along *y* or *z* for all porosities. Clearly, phonons in *x* encounter a dense 'surface' of voids, which reduces the *line-of-sight*, and introduces large thermal resistance, by not allowing phonons enough space to get through. Consequently, phonons propagating perpendicular to the pores along *y* (rhombus symbol), have higher thermal conductivity compared to the *x*-direction (square symbol). Along the *z*-direction (into the page, hexagon symbols), the conductivity is the highest. Although here the phonons travel along surfaces which still have a significant degrading influence, they do not encounter any direct scattering along their transport direction. Thus, clearly, although surfaces can slow down phonons, the most important detrimental effect in their propagation in nanoporous materials is the reduction in their mean-free-path by reducing their *line-of-sight* — in other words, obstructing phonon propagation.

Influence of staggering: To further investigate the importance of *line-of-sight* on phonon transport, we placed pores and voids such that they are offset, and compared the results to a similar system where the pores/voids are uniformly aligned (see system schematics in Fig. 5a). Voids were placed such that they were offset perpendicular to *x,* i.e. offset in *y* and *z* (see Fig. 5a (ii)). We find that staggering in the positions of the voids does not have a significant influence in the thermal conductivity (see blue and orange data points in Fig. 5b for $\phi = 0.5\%$ and $\phi = 5\%$, respectively). This is because in the case of voids the phonons can propagate around them more easily. The discrepancy between uniformly aligned and offset scatterers, however, has a stronger effect on systems with pores, which restrict phonons to flow around them only in the *xy*-plane. For a low 2% porosity, relocating pores as to reduce the *line-of-sight* for phonons travelling in the *x*-direction, results in a decrease of ~ 25% in the thermal conductivity (see magenta systems and data points in Fig. 5). The effect of *line-of-sight* reduction in staggered geometries compared to uniform geometries is illustrated in Fig. 5c). Clearly the phonon pathways and areas in which they can flow uninterruptedly are reduced in the staggered geometry.

We point out that to lower the uncertainty of this estimate, both systems being compared were simulated for sets of 20 simulations. At higher porosity (5%) the effect of staggering on the thermal conductivity is seemingly less marked, with only a 15% difference, which is however within the statistical error of our simulations (see the grey data for $\phi = 5\%$). This is to be expected, because at higher porosities, the average mean-free-path has already been significantly reduced by pores/void scattering, in which case phonon trajectories are randomized more similarly in both the aligned and staggered cases.

To investigate the influence of phonon relaxation times on the staggered (versus aligned) pore geometries, in the inset in Fig. 5b we show the evolution of the cumulative integral of the HCACF, from which the thermal conductivity is calculated. The integral of the HCACF is equivalent early on until around 10 ps for the corresponding staggered and aligned systems. After that point they diverge: the integral of the staggered systems plateaus (dashed lines), while that of the aligned systems continues to grow (solid lines). This indicates that both aligned and staggered systems undergo similar relaxation phonon processes for fast relaxing phonons (i.e. phonons which thermalize fast - those with short mean-free-paths). The difference in thermal conductivity arises due to slow relaxation time processes (long mean-free-path phonons) that are only still present in the aligned systems. This suggests that it's the suppression of larger mean-free-path phonons that affects the thermal conductivity in the staggered pores geometries. The short MFP phonons thermalize and are randomized before they meet the next scatterer, whereas the long MFP phonons travel in-between the pores in the aligned case, but scatter straight on the pores when they are staggered. Our observations agree with other works in the literature, at similar porosities as well [22]. As a side note, we have also computed the thermal conductivity in the $z$-direction, parallel to the pores for the lower porosity system in magenta. There, for an unexpected reason, staggered pore geometries had a lower thermal conductivity as well, by 21%. This is within the statistical error of the simulations, otherwise this would suggest that staggering affects transport perpendicular to the pores as well.

<u>Influence of pore clustering:</u> To continue towards building our understanding of disorder in nanoporous materials, we proceed by performing simulations in systems that contain clusters of voids/pores, either aligned or misaligned, as one might encounter in a realistic material. In order to clearly demonstrate the effect of clustering on $\kappa$, and get a quantitative understanding of its importance, we first begin by comparing the thermal conductivity of two systems with equivalent porosity, but different void arrangements as shown in Fig. 6a, systems (i) and (ii). System (i) in Fig. 6a exhibits clustering perpendicular to the transport direction, i.e. $x$, similar to systems discussed so far. System (ii) in Fig. 6a has equivalent porosity, but the distance between voids is equivalent in all directions such that the system is purely isotropic. The thermal conductivity of system (i) is ~35 % lower than that of system (ii). This is as expected, since system (i) has reduced the *line-of-sight* for phonons, by creating a *wall* or *barrier* of voids.

Next, we proceed with simulating the systems in Figs. 6b, where we consider the effect of clustering along the transport direction. In disordered nanoporous materials clustering happens in a statistical fashion, and increases local resistance, which reduces the overall $\kappa$. This is a well-defined effect that we have investigated using ray-tracing Monte Carlo simulations in the past for much larger pore sizes and geometries [24,28]. Here, as a first step, in Fig. 6b, we consider the effect of clustering on aligned systems, so as to isolate the effect from others at play in randomized systems. We start from a geometry of spread out voids (left column)/pores (right column), shown in Fig. 6b, and we then compress the positions of the voids/pores such that they are placed very closely together. We consider situations where the placement remains aligned, and where the placement is staggered.

Fig. 6c shows the first 50 picoseconds of the HCACF integral for the systems shown in Fig. 6b (with the same coloring identification). The converged thermal conductivity values (extracted at 150 ps) are also indicated in the right panel of Fig. 6c as a function of porosity for pores and voids alike. When it comes to phonon scattering on voids/pores, multiple scattering events are needed to fully thermalize phonons. For this reason, larger cluster sizes are more effective in thermalizing phonons [45]. On the other hand, spread out pores (or voids) are highly effective in scattering phonons with mean-free-paths equal and greater than the distance between them, but isolated pores are less effective scatterers than the clusters. Both effects work against each other, and no further degradation in the thermal conductivity of the void geometries is observed in our simulations by clustering defects. In fact, looking at the thermal conductivity (in Fig. 6c) of the void systems of Fig. 6b, left column, we observe that the spread-out system (orange), has a lower thermal conductivity than the aligned clustered system (brown). It is only when the voids are staggered (blue system), rather than aligned, in which case the *line-of-sight* is somewhat reduced, that the thermal conductivity of the clustered system is reduced. Still, however, only to the levels of the spread-out pore system (orange) (i.e. the results are within each other's error bars). Thus, in the void systems we consider, both clustering and staggering have a negligible effect on the final thermal conductivity (at fixed porosity). This is mainly because the *line-of-sight* reduction is not strong, owing it to the phonons being able to propagate around the voids.

We consider now the systems with pores (in magenta, orange, cyan and green) in the right column of Fig. 6b. Again, as in the case of voids, while clustering still keeps the pores aligned, it produces similar thermal conductivity reductions as having the pores spread uniformly along the channel transport length (compare the cyan and orange to the magenta-spread out systems/lines) at 50 ps in Fig. 6c and in the right panel of Fig. 6c (at 150 ps). The variations in thermal conductivity between these three systems are within each other's statistical error and not remarkably different. Staggering the clustered pores (green system), however, yields a significant ~ 43% decrease in thermal conductivity compared to the uniformly distributed case (magenta — right column of Fig. 6b). Thus, we reach here the conclusion that beyond defect scattering, the effect of reducing the *line-of-sight* due to staggering (or equivalently randomly placing pores in a more realistic experimental scenario) is more important in lowering the thermal conductivity. The quantitative distinction between the larger reductions (and consequently *line-of-sight* reductions) for staggered pores compared to staggered voids, is geometrical — pores introduce a barrier in the *z*-direction (out of page) and allows only 2D passages around them, whereas voids still allow 3D passages around them.

An interesting point arises when considering the trends of the cumulative integral of the HCACF (see left-hand plot in Fig 6c). We can observe that for both voids and pores, in the uniform systems the cumulative integral monotonically increases (lower orange line and magenta line). (The void systems have lower thermal conductivity because of their higher porosity.) On the other hand, the clustered systems plateau faster, after an initial sharp increase in the HCACF integral. We can infer from this result that slow processes (i.e. large mean-free-paths) thermalize much faster in the clustered geometries compared to the corresponding aligned geometries. Placing voids/pores at regular intervals contributes to annihilating larger mean-free-path phonons, whereas by spreading out the pores/voids we hinder mid- to short-MFP phonons more effectively. In the case of staggered systems of voids and pores (geometries in dark blue and green in Fig. 6b, respectively), the HCACF behaviour is initially similar to the equivalent clustered systems (in brown for voids, and orange or cyan for pores), with the exception that there is a marked *dip* in the autocorrelation function after ~10 ps, corresponding to slightly anti-correlated behaviour in the heat-flux, and which reduces the thermal conductivity. This interesting effect is a characteristic of some liquid and amorphous materials [50-52], and could possibly be due to oscillatory behaviour in the HCACF resulting from ballistic phonons moving back and forth [53]. It is interesting that our results suggest that voids/pores materials exhibit anticorrelated heat-flux behaviour, suggesting liquid, amorphous, or oscillatory behaviour, which we will be investigating in future work. However, the basic result we want to outline from this study, is that even down to the nanoscale, the intuitive *line-of-sight* argument, comes to be an important (if not the most important) feature in understanding thermal conductivity in pore/void filled materials.

<u>Influence of void size and position randomization:</u> Finally, in order to address the degree of disorder that one would encounter in a realistic system with voids, we consider void systems with random positions and sizes. The anticipation is that the effects we described above, clustering, surface area *vs.* volume, and *line-of-sight* reduction, all will simultaneously be present in disordered systems. In Fig. 7 the following geometries are compared: (i) uniform distributions (at $\phi = 5\%$, $\phi = 10\%$, and $\phi = 15\%$), (ii) random position distributions with a fixed void radius of 1.56 nm (at $\phi = 5\%$ porosity) (iii) random position distributions with a fixed void radius of 0.5 nm (at $\phi = 5\%$, and $\phi = 15\%$), and (iv) random position distributions with a randomly distributed void size based on a normal distribution with a mean, $\mu = 1$ nm, and a standard deviation, $\sigma = 0.2$ nm (at $\phi = 5\%$, $\phi = 10\%$, and $\phi = 15\%$). Three sets of realizations were simulated for geometries (ii)–(iv). Looking at the 5 % porosity, we find that randomizing void distribution (system type (ii)) while maintaining void size, yields only a small additional reduction (4–12%, within error bars) in thermal conductivity compared to the aligned structure — see blue data points in Fig. 7b. This is consistent with results found for staggered and clustered systems, both of which seem to preferentially reduce large phonon mean-free-paths. The randomly distributed and sized systems, labelled (iv) in Fig. 7a and colored in green, show a greater reduction in thermal conductivity (29–40%). However, randomly distributed systems of smaller nanovoids, fixed at ~0.5 nm (i.e. like the light-blue structure in Fig. 7, labelled (iii) in Fig. 7a), show an even greater reduction in thermal conductivity (69–71%). Several effects could contribute to this: smaller voids increase surface area (and thus scattering), but they also increase the total number of scatterers. Thus, there is a balance between sparsity (sparse scatters reduce the average phonon mean-free-path) and clustering (which more effectively thermalizes phonons).

In addition, we observe that the effect of having more (and smaller) voids on lowering thermal conductivity seems to increase with porosity. We see this by comparing the overall reduction in thermal conductivity of randomized geometries compared to uniform distributions as the porosity increases. This can be seen by considering how sets (iii) and (iv), in light-blue and green, respectively, in Fig. 7b decrease thermal conductivity more effectively as porosity increases, compared to the respective pristine systems. A more detailed plot including the percentage decrease in thermal conductivity for each set is shown in the *Supplemental Material*. Regarding the effect of void size, smaller voids do not become more influential at higher porosities (comparing systems (iii) and (iv), again in light-blue and green in Fig. 7). At $\phi = 5\%$ and 15% porosities, we can see that reducing the size of the voids (iii) decreases the thermal conductivity an additional 29–42% at $\phi = 5$ %, but only an additional 20-23% at $\phi = 15\%$). At higher porosities, the average phonon mean-free-path has already been significantly reduced (overall lower thermal conductivity), which could explain why reducing void size can only change the thermal conductivity so much (from 60–63% to 83% at 15% porosity compared to a change from 29-40% to 69-71% at 5% porosity). Similarly, the reduced spread in the data for the three systems (in light blue) at 15% porosity (compared to 5% porosity), suggests that not only the minutia of the size distribution has a weaker effect at higher porosities, so does the void position distribution. That said, our data suggests that in absolute terms, smaller randomized voids can still be effective in scattering shorter mean-free-path phonons (at higher porosities).

<u>Mean-free-path analysis:</u> Lastly, we compute the phonon thermal conductivity accumulation function, $k_{accum}$, for the pristine Si material using lattice dynamics, and perform a simple resistive MFP analysis based on Matthiessen's rule to estimate the effect of each MFP in the case of the nanostructures. This will allow us to better connect the actual geometrical features with the phonon MFPs. This calculation is performed using lattice dynamics and the phonon Boltzmann Transport Equation (BTE) solved under a relaxation time approximation (RTA) as implemented in the open-source package ALAMODE [54]. A 2 x 2 x 2 supercell (of 64 atoms) is used for all of the phonon and interatomic force constant (IFC) calculations. The selected atomic displacements are of 0.01 and 0.04 Å for the harmonic and anharmonic IFCs, respectively, and cubic interaction pairs up to the second nearest neighbour are considered." The Brillouin zone is sampled with a 30 x 30 x 30 q-point mesh for all calculations. We perform this for the perfectly crystalline pristine system for the Stillinger-Weber (SW) potential we use.

The normalized thermal accumulation function, $k_{accum}/k_{bulk}$, for the SW potential calculated at 300K (black line) is shown in Fig. 8a. For comparison, we also show the $k_{accum}/k_{bulk}$ computed using first principles, with a 18 x 18 x 18 q-point grid at 277K, by Esfarjani *et.* al [55] (blue line). The shape of the two functions is very similar, however, the SW function is shifted towards higher MFPs, a consequence of the higher thermal conductivity that this potential provides. In spite of this difference, the normalized thermal conductivity accumulation follows a distribution that is universal in shape for the same type of materials, regardless of the approach used to compute it [56-59]. (The value for the $k_{bulk}$, given by the SW potential using the BTE (with ALAMODE) is 592.85 W/mK, ~4 times larger compared to the value obtained with first principles. This is well known for this potential, and in agreement with other works [55]. This would mean that defects will have a larger effect in the thermal conductivity computed using SW rather than other potentials, but qualitatively, we can still draw conclusions on the influence of defects on MFPs.

To estimate the effect of each MFP in the case of the nanostructures, as it is computationally very expensive to perform a similar analysis for the larger nanostructures, we performed a simple resistive MFP analysis based on Matthiessen's rule as follows:

1. We begin by computing the contribution of $\Lambda_i$ to the thermal conductivity of the pristine system, $k_{pristine,i}$ by differentiating $k_{accum}$. This is shown in the inset of Fig. 8a. (The equivalent plot of the inset in Fig. 8a for the first principles calculation shown in Fig. 8a (in blue) can be found in the same Ref. [55]).

2. We then use Matthiessen's rule to combine the pore scattering MFP to each pristine MFP, $\Lambda_i$, and thus evaluate the contribution of the combined MFP to the thermal conductivity (in the $x$-direction) of a porous geometry, $k_{porous,i}$, as:

$$\frac{1}{k_{porous,i}} = \frac{\Lambda_i}{k_{pristine,i}} \left( \frac{1}{\Lambda_i} + \frac{C}{d} \right), \tag{3}$$

where $k_{pristine,i}$ is the contribution of $\Lambda_i$ to the thermal conductivity of the pristine system, as aforementioned. Above, $d$ is the distance between pores along the transport direction, $x$, and $C$ is a scattering strength measure of the defects encountered every $d$. In effect, $C$ is determined by the details of the pore arrangement within each "clustered wall" of pores, and is linked to local porosity and disorder (the larger the local porosity and disorder, the larger the local resistance, and the larger $C$ is).

3. We adjust the scattering strength/resistance parameter, $C$, to map to the thermal conductivity MD results for some example nanostructures. I.e. we find $C$ for each individual structure such that we get the MD calculated thermal conductivity at the end of the accumulation function. The corresponding accumulation functions are shown in Fig. 8a, for characteristic structures (some of are shown in Fig. 9(b)). This is a first order indication of how each MFP is affected in the case of the nanostructures. The colors in Fig. 8b are chosen to match those of the uniform geometries in the *Supplemental Material* as well (Table S1).

Note that, for the pristine geometry, even at 10 μm, $k_{accum}$ is lower than $k_{bulk}$. For this reason, the projected values of $k_{accum}$ for the nanostructures (i.e. the points plotted on the right side of Fig.8b) are normalized to $k_{accum}$ at 10 μm, i.e. the thermal conductivity of the pristine Si obtained with MD is matched to $k_{bulk}$. We further remark here that we obtain the same results using a smooth, interpolated, $k_{accum}$ (and thus $k_{pristine}$) as we do with the actual $k_{accum}$, despite its step function shape at high MFPs, which translates in the spikes in the inset of Fig. 8a.

Next, we observe how the *C* parameters relate to the geometrical details of the defected areas that allow phonon propagation. In Fig. 9(a) we plot the *C* parameter (which provides the strength of the scattering defect 'line') versus the area of the defects in uniform nanostructured families indicated with the star-symbols (i.e. the local porosity, in the vicinity of the pore regions), showing very good correlation. A linear dependence is observed with $D^2$, such that $C = sD^2$, where *s* is the slope of the fit and *D* is the pore diameter. For example, the structures in magenta and grey (Fig. 9b(i)), have their defect 'walls' at the same distance *d*, but the pore sizes increase from 0.87 to 1.37 nm (thus reducing the region available for phonon propagation), and that is clearly reflected on the *C* value, which is approximately halved. Both geometries fall closely near the linear fit, as do geometries with different *d*, and similar *D* (see the green and magenta geometries in Fig. 9b(iii), for example). It is quite interesting that this linear fit is achieved with only one parameter *C* for each structure, which can very well also be linked back to the underlying geometry, and specifically to the geometrical details of the defected (i.e. porous) areas that allow phonon propagation.

In the case of staggering, where the *line-of-sight* is further reduced, to match the $k_{accum}$ function, the *C* parameter must be higher (shown by the colored magenta and grey circles), clearly indicating the increase in the effective resistance of the "defect wall" that the phonons encounter. Notice that the magenta and grey circles correspond to the same color stars, which are the equivalent, non-staggered systems. The colors in the above figure are again chosen to match those of the geometries in the *Supplemental Material* (this includes all geometries in Table S1 (stars) and the pore geometries in Table S4 (circles)).

In one final illustration, we use the extracted *s* from the linear fit in Fig. 9a ($C = sD^2$) to compute $k_{accum}$ for the nanostructures (which gives a slightly different *C* in each case). In Fig. (c), the simulated conductivity of the nanostructures is shown along with their 95% confidence interval from MD (error bars) on the right, using the same colors. Here, *s* is found to be 0.105 nm$^{-2}$. This is just to show that quite accurate predictions can also be obtained for other similar nanostructures as well, in correlation to simple geometrical details. We also find that the void systems exhibit a similar relationship and we include the analysis of the uniform void geometries in the *Supplemental Material*. In general, one can think of analysing the *C* or *s* values for structures with more complex topologies, and that will provide an indication of the strength of the relative local scattering/resistance that is introduced. This analysis, however, shows again how simple considerations, such as Matthiessen's rule and the particle picture can provide adequate understanding of heat transport even down to the nanoscale.

## IV. Conclusion

In conclusion, we used large-scale equilibrium molecular dynamics together with the Green-Kubo formalism to examine thermal transport in over 50 Si nanoporous/nanovoid geometries, incrementally varying their degree of disorder, from uniformly distributed, to staggered (or offset), clustered, clustered and staggered, randomly distributed, and randomly distributed and sized pores. Porosity, surface-area, pore size, neck size, and pore number have all been examined as well. Our goal was to clarify the effect of small, systematic variations in pore/void geometry on thermal conductivity in order to determine the most effective mechanisms in reducing thermal conductivity. We show that surface area and pore/void density is a more significant metric in estimating thermal conductivity reduction compared to porosity and pore sizes, with the exception when large pores are placed in such a way as to block the passage of phonons, an effect referred to as the reduced *line-of-sight*. It turns out that the most drastic reductions in thermal conductivity in nanoporous materials can be explained by this effect at first order. For transport perpendicular to pores, staggered pores reduce the *line-of-sight* more and are thus more effective in reducing the thermal conductivity compared to aligned pores, an effect that is more marked at lower porosities. Unlike staggered pores, staggered voids are not as effective in reducing the *line-of-sight*, as phonons can flow around them in 3D. We also show that the clustering of pores/voids itself does not seem to influence the overall thermal conductivity compared to an equivalent uniform system, unless the clusters are in a staggered geometry, in which case the *line-of-sight* is reduced. Clustering, however, contributes to annihilating larger mean-free-path phonons, something observed from the HCACF. In the case of fully randomized void geometries, in terms of void position and size, we find that having more (and smaller) randomized voids lowers the thermal conductivity compared to the uniform system. The strength of this effect seems to increase with porosity, in agreement with macroscale Monte Carlo works as well [24], although the details of the size and position distributions seem to have a weaker effect at high porosities. Finally, we also show that transport along the pores, i.e. parallel to them, is more susceptible to the distance between the pores, as they reduce the 'clean' regions for phonon flow. In that case, smaller but densely packed pores reduce thermal conductivity more effectively compared to larger but spread out pores. Ultimately, the basic result we want to outline from this study, is that even down to the nanoscale, the intuitive *line-of-sight* argument comes to be maybe the most important feature in understanding thermal conductivity in pore/void filled materials. As a consequence, we find that a simple model, based on Matthiessen's rule and simple geometrical considerations, yields thermal conductivity accumulation function results for (uniformly distributed pores and voids) nanostructured geometries which agree with the MD simulations. It is interesting that such an understanding is drawn from wave-based MD simulations.

## Acknowledgements

This work has received funding from the European Research Council (ERC) under the European Union's Horizon 2020 Research and Innovation Programme (Grant Agreement No. 678763). LdSO thanks the following people, all of whom provided useful insights towards this work: Dhritiman Chakraborty, Samuel Foster, Chathurangi Kumarasinghe, Patrizio Graziosi, John McCoy, Aria Hosseini, and Cidália Sousa.

Figure 1:

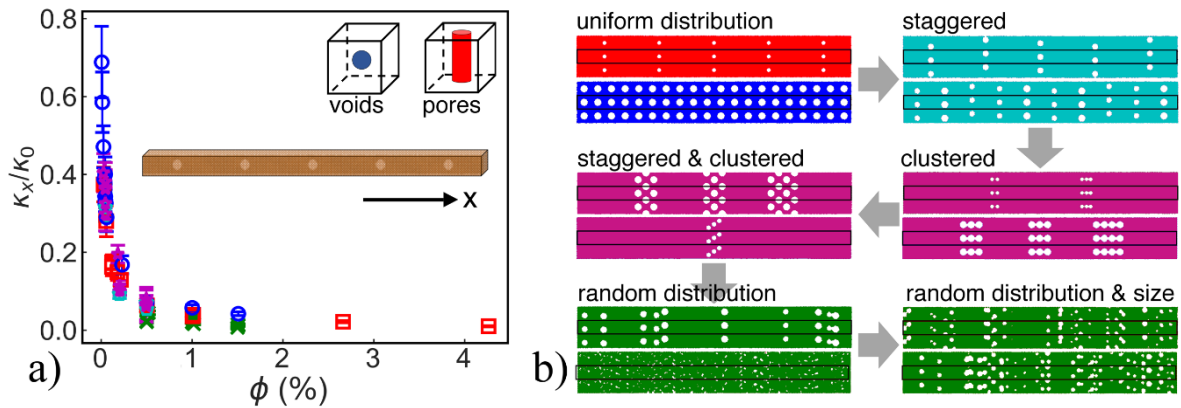

Figure 1 caption:

(a) The fractional thermal conductivity as a function of porosity, computed along the *x-axis* for the geometries considered in this study. A 3-dimensional representation of a simulation cell is included as an inset in (a). The color coding in (a) corresponds to that shown in the geometries in (b). For uniform distributions (top left in (b)), a distinction is made between systems with pores in blue (cylindrical holes) *vs.* voids in red (spherical holes). The examples of simulated systems shown in (b) are chosen to illustrate the variability in the geometries considered. Variations in the geometries include changes in the number of pores/voids and in the void/pore diameter (top left figures in red and blue), as well as staggered (or offset) voids/pores (top right figures in cyan), clusters of voids/pores both aligned and misaligned (center right and left figures in magenta, respectively), and geometries with randomly distributed voids (bottom left), and randomly distributed and sized voids (bottom right). The figures in (b) correspond to a cross section of the *xy*-plane, i.e. perpendicular to *z,* and include the periodic images along *y*.

Figure 2:

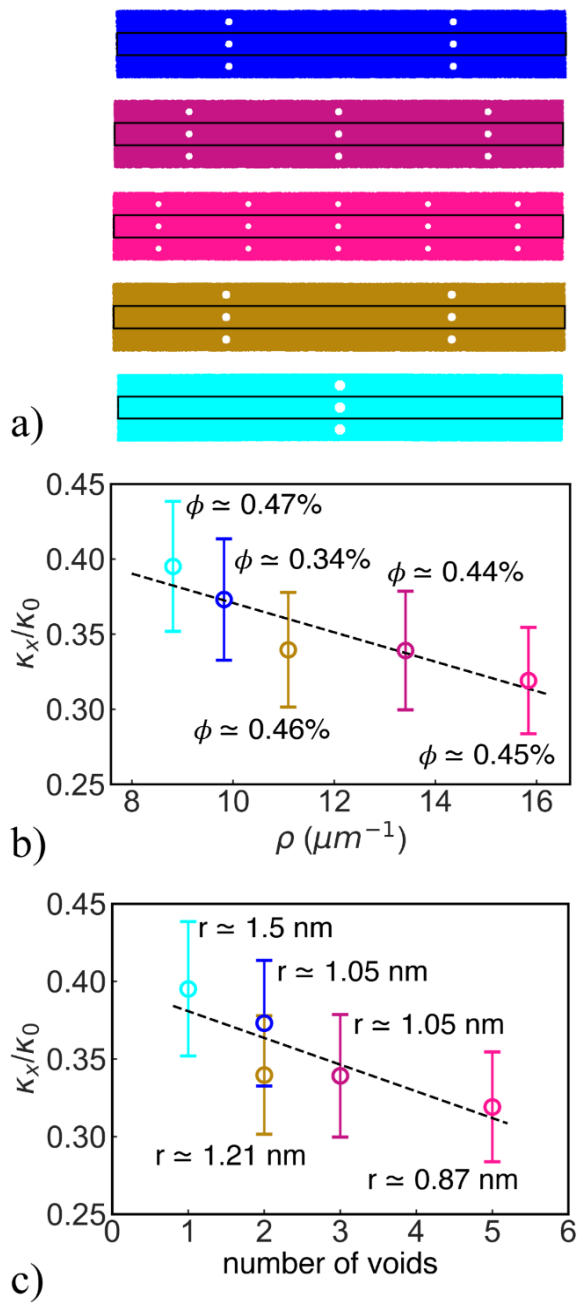

Figure 2 caption:

Effect of surface area and number of voids on thermal conductivity. The geometries considered are shown in (a). Four systems (in purple, magenta, brown, and cyan) have near equivalent porosities of $\phi \sim 0.45\%$, but different surface areas. We include a fifth (blue) system with lower porosity, $\phi = 0.34\%$, but with a surface area in the vicinity of the other systems' surface area. (b) The thermal conductivity of these geometries as a function of the surface-area-to-volume ratio, $\rho$. The porosities of the geometries are displayed next to each data point, and the colors in (b) match the colors of the actual geometries shown in (a). (c) The thermal conductivity versus the number of voids. The radii of the voids for each of the geometries is as indicated. In both (b) and (c), the black dashed lines show linear fits to the data.

Figure 3:

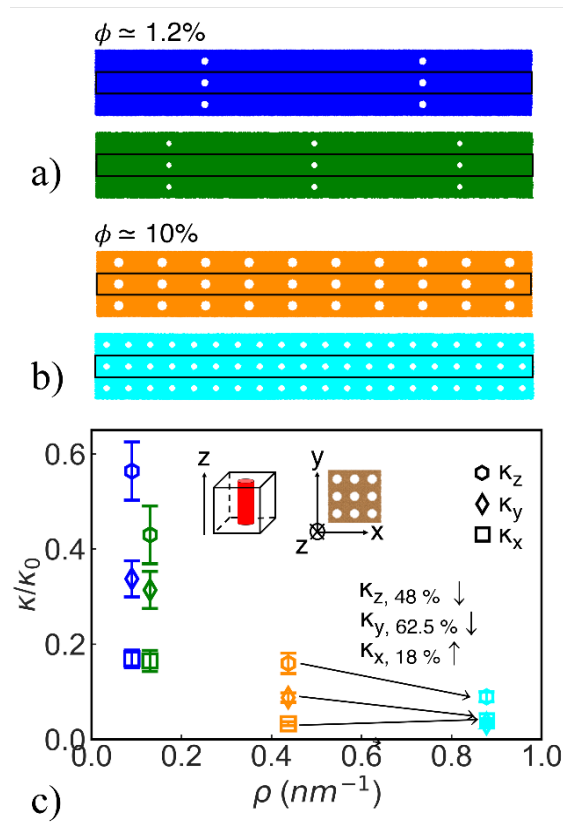

Figure 3 caption:

The effect of surface area on the thermal conductivity for porous geometries. The geometries considered are depicted in (a) and (b), with (a) small ~1.2 % and (b) large ~10.1 % porosities, respectively. (c) The thermal conductivity as a function of the surface-area-to-volume ratio for the two sets of geometries. The colors in (c) match the corresponding geometries in (a) and (b), whereas the different symbols indicate the thermal conductivity along the different cartesian directions of the simulation cell, *x*, *y*, and *z*. The anisotropy between $\kappa_x$ (square symbols), $\kappa_y$ (rhombus symbols) and $\kappa_z$ (hexagon symbols), is due to the anisotropy in the different phonon paths imposed by the pore locations. Considering all directions is revealing of the effect the distance between pores, in addition to surface area, exerts on thermal transport.

Figure 4:

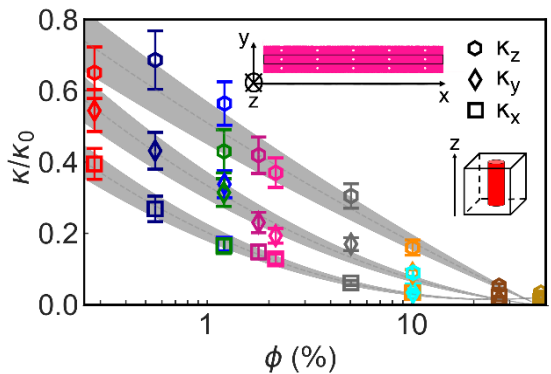

Figure 4 caption:

Asymmetric thermal transport in porous systems. The thermal conductivity of geometries with pores is evaluated for uniformly distributed pore systems in all three directions, i.e. in $x$ (square symbols), $y$ (rhombus symbols), and $z$ (hexagon symbols). Porosity, along each cartesian direction ($z$, $y$, and $x$, from top to bottom) is plotted in logarithmic scale, and the shaded regions highlight fits to the corresponding error bars, while the dashed-grey lines show fits through the data. The error bars show the standard error of all simulations in a given geometry set. At equivalent porosity, pores offer greater resistance when transport is perpendicular (i.e. in the $x$- and $y$-direction) to them. The relative gap in thermal conductivity between transport along $y$ and $z$ increases as porosity increases. The insets show (on the upper/right) an overview of a cross section of a simulation cell for an indicative geometry (the simulation domain is indicated by the black outline, and periodic boundary images along $y$ are also shown). A representation of a pore geometry (only pores are considered in this figure) is also represented as an inset (lower/right). The directions we consider are labelled accordingly.

Figure 5:

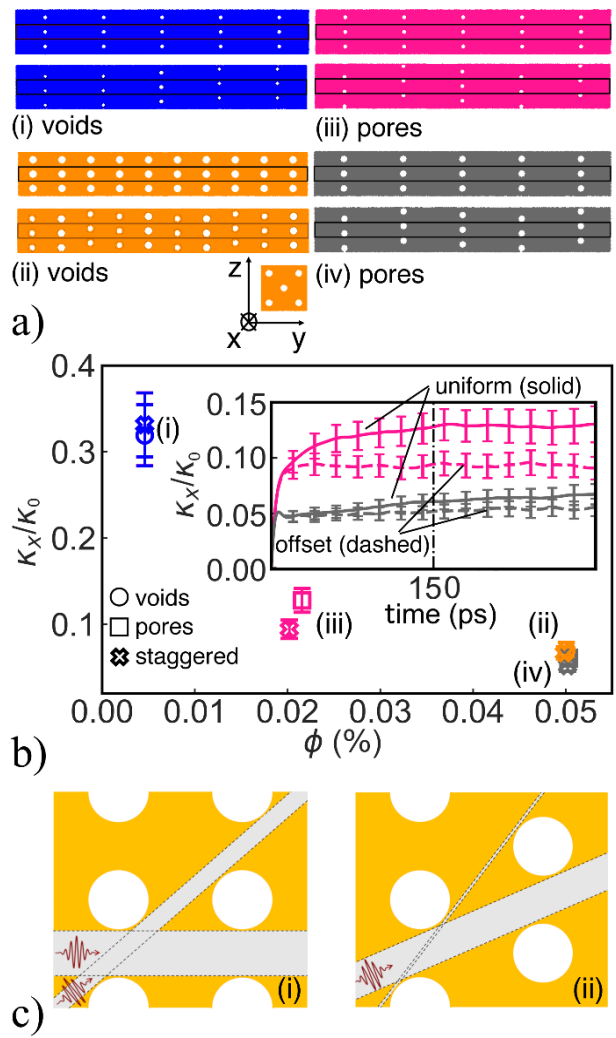

Figure 5 caption:

The effect of staggering (offset placement of voids/pores) on the thermal conductivity. In (a) systems with uniformly distributed voids (i and ii) and pores (iii and iv) are placed so they are offset to each other as shown: voids are offset in the *y*- and *z*-directions, whereas pores are only misaligned the *y*-direction. In this way, phonons propagating along *x* will 'feel' the misalignment, but not when propagating along *y*. (b) The thermal conductivity of these geometries — for comparisons between staggered pores and voids geometries versus aligned pore/void systems. The colors in the plot match the geometries in (a), with aligned voids labelled by circles, aligned pores labelled by squares, and offset pores and voids both labelled by crosses. The inset in (b) shows the evolution of the thermal conductivity along the HCACF time for the pore geometries. The dashed lines correspond to the staggered geometries, and the solid lines to the aligned geometries. The thermal conductivity values in the main plot of (b) for the pores (magenta and grey) correspond to the HCACF value at 150 ps in the inset. (c) Illustration of phonon *line-of-*sight in (i) aligned and (ii) staggered pores. In both cases, the shaded regions show examples where the propagation of phonons is unimpeded, i.e. the line-of-sight of phonons. In the staggered systems (ii), the grey lines become thinner than in the aligned pore system of (i). In other words, the aligned pores have a wider range of unimpeded regions that allow phonons to propagate.

Figure 6:

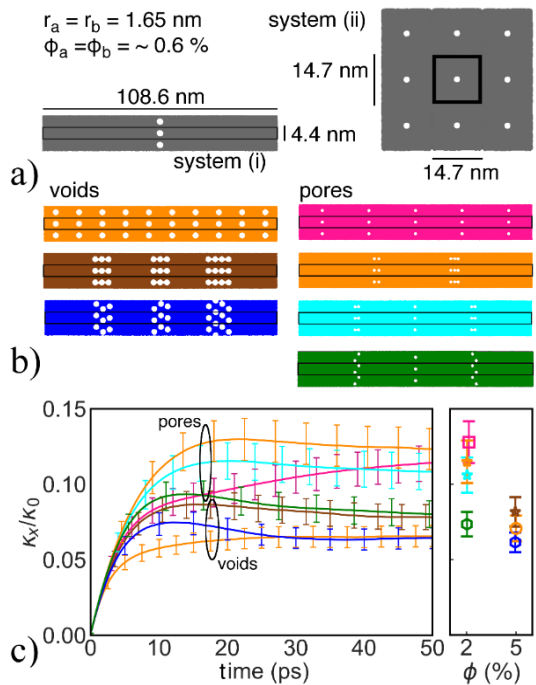

Figure 6 caption:

Effect of pore/void clustering on thermal conductivity in anisotropic geometries. (a) Schematic of wo systems with equivalent void size and system volume (and thus porosity). The thermal conductivity of system (i) in the *x*-direction, which corresponds to a clustering of the voids perpendicular to transport (due to periodic boundary conditions in *y,* and *z*), is ~35 % lower than that of system (ii), where the voids are truly uniformly distributed. (b) Simulated geometries of aligned systems for both voids and pores, and equivalent porosity and void/pore size geometries with arrays of clusters and offset arrays of clusters. (c) Left panel: The cumulative thermal conductivity as it evolves in HCACF time up to 50 ps. Right panel: The estimated converged thermal conductivity extracted at the 150 ps cut-off for the clustered and aligned systems of pores and voids in (b), color coded accordingly.

Figure 7:

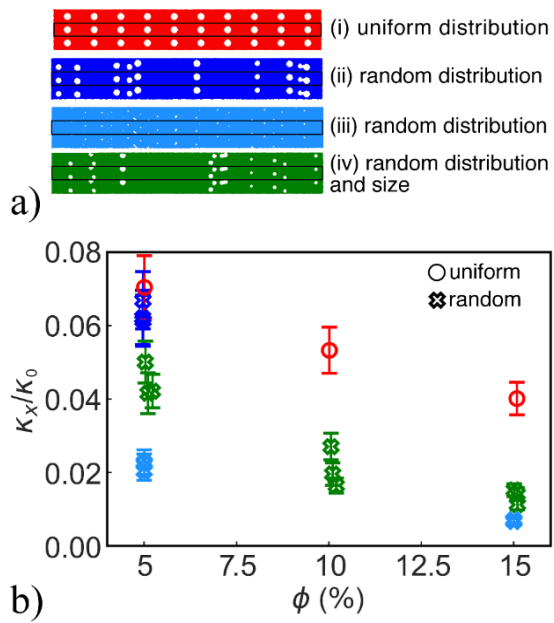

a)

b)

Figure 7 caption:

(a) Schematic of randomized types of geometries: baseline uniformly distributed voids (i), randomly distributed voids with fixed radius at 1.56 nm (ii), randomly distributed voids with fixed radius at 0.5 nm (iii), and randomly distributed voids with a randomly distributed void size based on a normal distribution with a mean of 1 nm and a standard deviation of 0.2 nm. Three realizations of each of the indicative random systems in (a) were simulated (the schematic in (a) are for 5% porosity). (b) Fractional thermal conductivity of randomized void systems at different porosity. The geometry types are colored according to the schematic in (a). Circles (in red) show the thermal conductivity for uniform geometries at different porosities, and the random systems are represented with crosses. Notice that for type (iii) simulations at 15% porosity, the variability between the three sets of simulated systems is so small that all thermal conductivities overlap and only a single (light blue) point appears visible, when in fact there are three.

Figure 8:

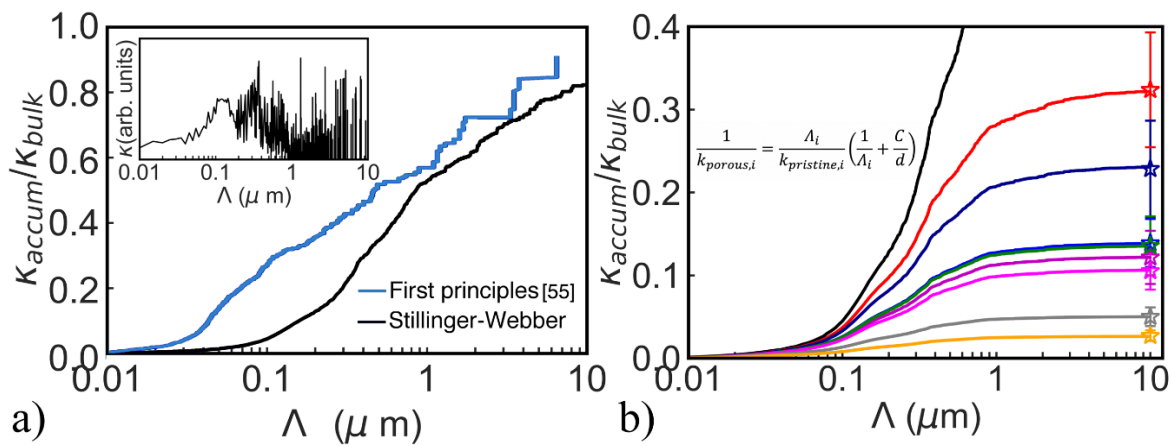

Figure 8 caption:

(a) Normalized thermal conductivity accumulation function computed for pristine Si using the Stillinger-Webber potential (black line), and from first principles (blue line) [55]. The inset in (a) shows the derivative of the Stillinger-Weber accumulation function, i.e. the contribution of each mean-free-path to the thermal conductivity. (b) Thermal conductivity accumulation functions for the uniform pore geometries. The functions were computed using the expression shown in (b) by finding $C$ such that $k_{porous,10\ \mu m}$ matched the thermal conductivity computed from molecular dynamics. The pristine geometry accumulation function is shown in black. The colors in (b) match those in Table S1 in the *Supplemental Material*.

Figure 9:

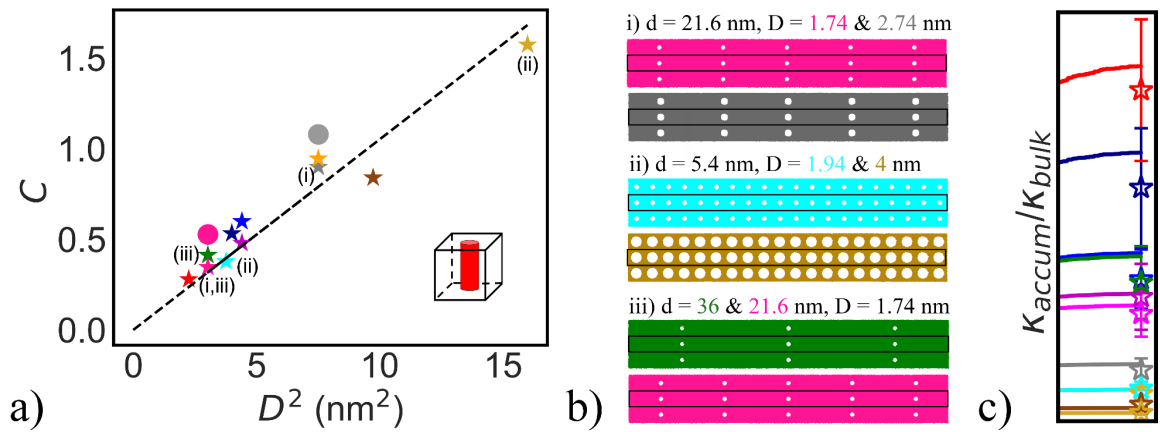

Figure 9 caption:

(a) Relationship between the scattering parameter $C$ and the square of the pore diameter, $D$. The data corresponds to uniformly distributed (stars) and staggered (circles) pore geometries. A linear fit through the data (black line), which is forced to go through the origin, is also plotted. (b) Examples of some of the geometries included in (a) and their corresponding $d$ and $D$. (c) Expected value for the thermal conductivity accumulation function of the nanostructures computed for $C = sD^2$, for the uniform pore geometries. All results are within the 95% confidence interval of the simulated thermal conductivities (star symbols and corresponding error bars).